\documentclass[letterpaper, 10 pt, conference]{ieeeconf}  
\usepackage{amssymb}
\usepackage{mathtools}
\usepackage{amsmath}
\usepackage{verbatim}
\usepackage{soul}
\usepackage{color}
\usepackage{ifthen}
\usepackage{xkeyval}
\usepackage{xcolor}
\usepackage{tikz}
\usepackage{calc}
\usepackage{graphicx}
\usepackage{fancyheadings}
\usepackage{booktabs}
\usepackage{multirow}
\usepackage{cite}
\IEEEoverridecommandlockouts                              

\newtheorem{theorem}{Theorem}

\newtheorem{assumption}{Assumption}

\newtheorem{remark}{Remark}

\newtheorem{lemma}{Lemma}
\title{\LARGE \bf Minimum-Energy Distributed Consensus of Uncertain Agents}

\author{Mohammad Zamani$^{1}$ \and Iman Shames$^{2}$ \and Valery Ugrinovskii$^{1}$
\thanks{*This work was supported by the Australian Research Council under Discovery Projects funding scheme ( project D0120102152 ) and a McKenzie Fellowship.}
\thanks{$^{1}$M. Zamani and V. Ugrinovskii are with School of Engineering and IT, 
        UNSW Canberra, Australia.$^{2}$I. Shames is with the Department of Electrical and Electronic Engineering, 
        University of Melbourne, Australia
        {\tt\footnotesize m.zamani@adfa.edu.au}, {\tt\footnotesize v.ougrinovski@adfa.edu.au}, {\tt\footnotesize  iman.shames@unimelb.edu.au}.}%
}

\begin{document}
\maketitle
\begin{abstract}
 This paper presents a consensus algorithm for a multi-agent system where
 each agent has access to its imperfect own state and neighboring state
 measurements. The measurements are subject to deterministic disturbances
 and the proposed algorithm  provides a minimum-energy estimate of the
 measured states which is instrumental in achieving consensus by the
 nodes. It is shown that the proposed consensus algorithm converges
 exponentially in the absence of disturbances, and its performance under
 bounded continuous disturbances is investigated as well. 
 The convergence performance
 of the proposed method is further studied using simulations where we show
 that consensus is achieved despite using large measurement errors.  
\end{abstract}
\section{Introduction}
 While there is a vast body of literature studying consensus in
 interconnected systems (e.g.~ see \cite{SaberCons07, bullo2009distributed,
   mesbahi2010graph} and the references there-in), the problem of reaching
 consensus in the presence of disturbances is not fully addressed. Here,
 the word \emph{disturbance} is used as a catch all phrase for noise,
 system parameter uncertainties, and computation errors. The existing
 results predominantly make assumptions on the nature of the disturbances
 (also called noise or uncertainty in the system). For example,
 \cite{huang2010stochastic} assumes that the noise signal are random, with
 a probability distribution, \cite{li2013distributed} assumes that model
 uncertainties are bounded by a known function, and
 \cite{bauso2009consensus} assumes that the disturbances are continuous
 over time. In the derivation of the algorithm in this paper we make no use
 of any knowledge about the nature or properties of the disturbances. 
 
 In this paper, we consider the problem of achieving consensus in a
 multi-agent network where the measurements of each agent from its own
 state and the states of the neighboring agents are subject to
 deterministic disturbances. We invoke a least squares deterministic
 filtering approach, namely the minimum-energy filtering that was pioneered
 by Mortensen~\cite{Mortensen} and later extended by
 Hijab~\cite{Hijab}. This method, requires no stochastic assumptions on the
 disturbance signals and has proven to be instrumental in non-standard
 filtering problems such as nonlinear filtering, cf.~\cite{AguiarTAC2006},
 and geometric filtering, cf.~\cite{ZamaniTAC2011}, due to its systematic
 least squares nature.  In this work, a minimum-energy filter is proposed
 for each node that provides distributed estimates of the states of that
 node and the nodes in the neighborhood by taking into account various
 errors of the model and the measurements. These errors include the state
 measurement disturbances, the initialization error of the node and the
 actuation (or quantization) error of the consensus algorithm. An important
 design innovation of the paper lies in calculating estimates of the
 neighboring states at each node that are not necessarily equal to the
 estimates that those nodes have calculated for their own state. Therefore,
 design conditions are provided that yield decoupled filter equations that
 are straightforward to implement in a distributed fashion and are scalable
 to large networks. The convergence of the proposed overall algorithm  for
 the disturbance-free case is shown using appropriate conditions for the
 proposed algorithm.  Also, we establish performance of the algorithm
 in the presence of disturbances, namely under bounded and continuous
 disturbances. 
The convergence performance of the proposed
 algorithm is also demonstrated via simulations which further support the
 aforementioned qualities of the proposed method. 
 
 The remainder of the paper is organized as follows. The formal definition of the consensus problem considered is provided in Section~\ref{consensus}. The proposed decoupled design methodology is introduced in Section~\ref{decoupled}. In Section~\ref{ME} we describe the filtering part of the problem that we tackle using the method of minimum-energy filtering.  
 The proposed filter and the details of its derivation are given in Section~\ref{filterder}. In Section~\ref{convergence} we summarize the overall proposed consensus algorithm and study its convergence properties in the case of scalar agent states. Section~\ref{numexp} contains a numerical example that demonstrates the convergence performance of the proposed algorithm using simulations. Finally,  Section~\ref{conc} concludes the paper. 
\section{Distributed Consensus Formulation}\label{consensus}
 Consider a network of $N$ agents with a directed graph topology $\mathbf{G}=(\mathbf{V},\mathbf{E})$ where $\mathbf{V}$ and $\mathbf{E}$ are the set of vertices and the set of edges, respectively. An edge directing node $j$ of the graph $\mathbf{G}$ towards node $i$ where $i,j\in\{1,\cdots,N\}$ is denoted by $(i,j)$. The network $\mathbf{G}$ is assumed to have no self-loops, i.e., $(i,i)\notin \mathbf{E}$. The neighborhood of node $i$, i.e., the nodes that node $i$ can obtain information from are denoted by $\mathcal{N}_i=\{j:(i,j)\in\mathbf{E}\}$. The Laplacian matrix of the network is defined as 
\begin{equation}
\begin{split}
&L\triangleq [L_{ij}],\quad L_{ii}=-d_i=-\sum_{j\in\mathcal{N}_i} a_{ij} \\
&L_{ij}=a_{ij} \mbox{ if } (i,j)\in\mathbf{E},\quad L_{ij}=0 \mbox{ if } (i,j)\notin\mathbf{E},
\end{split}
\end{equation}
where $a_{ij}>0$ is the weight of edge $(i,j)$.
 The degree matrix is defined as
\begin{equation}
\Delta \triangleq \mbox{Diag}[d_1,\cdots,d_N].
\end{equation}
The associated adjacency matrix of the network is then defined as 
\begin{equation}
A\triangleq  L + \Delta.
\end{equation}

 We consider a problem where all the agents of $\mathbf{G}$ are to reach
 consensus by utilizing imperfect measurements of their own states along
 with state measurements of the agents in their respective local
 neighborhoods $\mathcal{N}_i$. Consider the local dynamics 
\begin{equation}\label{state.dynamics}
 \dot{x}_i = u_i + B_i\delta_i,\quad x_i(0)= x_{i0},
\end{equation}
where $x_i\in\mathbb{R}$ is the state of agent $i$ initialized at
$x_{i0}$. The signal $u_i\in\mathbb{R}$ is the driving input that has to be
designed so that agent $i$ reaches consensus with all other agents. The
signal $\delta_i\in\mathbb{R}$ is a 
disturbance caused by numerical and/or actuation errors in the control
input $u_i$. The coefficient $B_i\in\mathbb{R}$ is known \emph{a priori}.   
A measurement of the local state, $y_{ii}\in\mathbb{R}$ is obtained according to the model  
\begin{equation}\label{measurementi}
y_{ii} = x_i + D_{ii}\epsilon_{ii},
\end{equation}
while measurements of the neighbouring states $y_{ij},\;j\in\mathcal{N}_i$ are obtained according to
\begin{equation}\label{measurementsj}
y_{ij} = x_j + D_{ij}\epsilon_{ij},
\end{equation}
where the coefficients $D_{ij}\in\mathbb{R}$ are known from the model, $D_{ij}\not = 0$. The
disturbances $\epsilon_{ij}\in\mathbb{R}$ are due to the errors of the
measurement instruments. Note that in the networked environment the errors $\{\epsilon_{ij}\}$ are
sometimes attributed to the communications errors between the agents. 

 Consensus using disturbance free state measurements is well studied,
 cf.~\cite{Olfati2005consensus}. The control signal in that case is often
 given by $u^*=\sum_{j\in\mathcal{N}_i}(x_j-x_i)$ where $\mathcal{N}_i$
 indicates the set of agents that are in the neighborhood of agent
 $i$. This allows agents to asymptotically reach consensus. In this paper however, we consider a case where one cannot use
 perfect state measurements but perturbed measurements of
type~\eqref{measurementi} and~\eqref{measurementsj}. Therefore, the
consensus objective must be revisited to account for disturbances. Namely,
in the absence of perturbations, we aim to achieve a standard state of average
consensus, but under disturbances we will aim to achieve an approximate
average consensus where the trajectories converge to a bounded
neighbourhood of the average consensus state.

To achieve the consensus objective described above, we propose using the
following local consensus method:
\begin{equation}\label{state}
\begin{split}
&\dot{x}_i =  u_i+ B_i\delta_i,\quad x_i(0)=x_{i0},\\
&u_i=\sum_{j\in\mathcal{N}_i}(\hat{x}_{ij}-\hat{x}_i).
\end{split}
\end{equation}
In equation~\eqref{state}, $x_i$ is the true state of agent $i$ and the
input $u_i$ is designed using  the estimated state of agent $i$,
$\hat{x}_i$, and using the estimated states $\hat{x}_{ij}\in \mathbb{R}$ of
the neighbours of agent $i$, $j\in\mathcal{N}_i$. The method to obtain
these estimated signals will be provided later in Sections~\ref{ME}
and~\ref{filterder}.  

 In the following we provide a distributed filtering algorithm to obtain
 the required local estimates $\hat{x}_i$ and $\{\hat{x}_{ij}\}$. The aim
 is to obtain these estimates, including estimates of the neighbouring
 agents $\{\hat{x}_{ij}\}$ locally at node $i$, as opposed to receiving
 them from the respective agents.  
\section{Decoupled Filtering Design}\label{decoupled}
Here, decoupled filtering design refers to a network of filters that have
decoupled dynamics. The consensus algorithm~\eqref{state} involves the
estimates $\hat{x}_i$ and $\hat{x}_{ij},\;j\in\mathcal{N}_i$. Our decoupled
design requires that node $i$ should calculate these estimates by measuring
its own state along with measuring the states of its neighbours. In
contrast, in a coupled design node $i$ must receive the estimates
$\hat{x}_{ij}$ from their respective neighboring agents, instead of
calculating them locally. Our interest in a decoupled design owes to the
fact that receiving 
$\hat{x}_{ij}$ with no communication errors is unrealistic and hence node
$i$ will need to use a filter to account for the communication errors of
the received information.   

 A key idea of this paper in order to circumvent these problems is to take advantage of the fact that all the agents of the network are to implement similar strategies with the shared goal of reaching consensus. We propose to approximate the neighboring state $x_{j}$ at node $i$ by
\begin{equation}\label{belief}
x_{ij}  = x_i + \eta_{ij},
\end{equation}    
 where 
 $\eta_{ij}$ is the approximation error. This formulation 
 is motivated by the fact that the neighboring node $j$ is running a
 similar consensus algorithm and therefore it makes sense for node $i$ to
 approximate the current state $x_j$ of its neighbour $j$ by its own state
 $x_i$ plus approximation error. This allows us to obtain a consensus
 algorithm at node $i$ that is decoupled and independent of the knowledge
 of the actual neighboring state $x_j$. This will be done in the next
 section where we will derive a distributed filtering algorithm using a
 minimum-energy filtering approach and by using the notions we have
 introduced so far.    

 \begin{remark}
A similar idea has been employed in~\cite{ZU_1} by the authors, where the H$_\infty$
distributed filtering in a network of filters subject to disturbances is
studied. In 
that work, a network of filters are required to provide local estimates of
the state of a reference plant by sharing information regarding their
estimates subject to disturbances. Apart from the difference between the
two problems, the difference to the idea employed here is that
in this paper 
each agent approximates the state of the
neighboring agents with its own state plus an
approximation error term; i.e. there is no reference for the agents to follow.   
 \end{remark}
\section{minimum-energy filtering}\label{ME} 
In the following a minimum-energy filtering problem is formulated that will yield the estimated states, $\hat{x}_i$ and $\hat{x}_{ij},\;j\in\mathcal{N}_i$, that are needed to drive the dynamics~\eqref{state} of agent $i$. 

 Minimum-energy filtering~\cite{Mortensen,Hijab} is a deterministic
 filtering approach that for linear systems results in the same Kalman
 filter~\cite{Kalman} equations, cf.~\cite{Willems}. Several arguments have
 been suggested in favor of minimum-energy filtering~\cite{Willems}. In
 this work, since we face a non-standard distributed filtering problem, we
 utilize minimum-energy filtering due its proven applications in complicated
 filtering problems such as nonlinear filtering, cf.~\cite{AguiarTAC2006},
 and geometric filtering, cf.~\cite{ZamaniTAC2011}.   
 
  In minimum-energy filtering, the error signals of the model are
  considered as unknown functions of time that are square integrable over
  any finite time interval $[0,t]$. A cost on the sum of square norms of these  error signals is considered,
\begin{equation}\label{cost}
\begin{split}
 &J_{i,t}(x_i(0),\delta_i,\epsilon_{ii},\{\epsilon_{ij}\},\{\eta_{ij}\}) =
 \frac{\Xi_i}{2} (x_i(0)-x_{i0} )^ 2\\
 &+ \frac{1}{2} \int_0^t [\delta_{i} ^2+   \epsilon_{ii} ^2  + \sum_{j\in\mathcal{N}_i}( \epsilon_{ij}^2 + \frac{\eta_{ij}^2}{G_{ij}})]d\tau.
 \end{split}
 \end{equation}  
Here, the variable $x_{i0}\in\mathbb{R}$ is an a priori candidate for the unknown initial state $x_i(0)$ and the scalar coefficient $\Xi_i\in\mathbb{R}$ is a weighting for measuring the energy of the initialization error $ \vert x_i(0)-x_{i0}\vert$. Similarly, the positive constant $G_{ij}$ is a weighting for the energy of the approximation error $\eta_{ij}$. The cost~\eqref{cost} is interpreted as the energy contained in the unknowns of the model~\eqref{state},~\eqref{measurementi},~\eqref{measurementsj} and~\eqref{belief} that is defined as the sum of the squared unknowns, the initial error $x_i(0)-x_{i0}$, the model error $\delta_i$, the measurement errors $\epsilon_{ii}$ and $\epsilon_{ij}$ and the approximation errors $\eta_{ij}$.

 The concept behind minimum-energy filtering~\cite{Mortensen} amounts to seeking a set of the unknowns $(x_i(0),\delta_i,\epsilon_{ii},\{\epsilon_{ij}\},\{\eta_{ij}\})$ consistent with the measurements $y_{ii}$ and $y_{ij}$ during the time period $[0,t]$ i.e., such that the model equations~\eqref{state},~\eqref{measurementi},~\eqref{measurementsj} and~\eqref{belief} are satisfied. A particular set $(x^*_i(0),\delta^*_i,\epsilon^*_{ii},\{\epsilon^*_{ij}\},\{\eta_{ij}^*\})$ that minimizes the cost~\eqref{cost} is selected since it has the  least energy content. Using this optimal set an optimal state trajectory $x^*_i$ and optimal approximate neighbor states $x^*_{ij},\;j\in\mathcal{N}_i$  are calculated for the period under consideration $[0,t]$. The last point of the optimal trajectory $x_i^*(t)$ and the instantaneous optimal approximate neighbor states $x^*_{ij}$  are then assigned as the minimum-energy estimates $\hat{x}_i(t)\triangleq x^*_i(t)$ and $\hat{x}_{ij}(t)\triangleq x^*_{ij}(t)$. As $t$ varies  this process is potentially a repetitive optimization process that needs to be redone for every period $[0,t]$ to yield the estimates at time $t$. In the following section however it is shown how to avoid the repetition by deriving a recursive filter that updates the estimates $\hat{x}_i(t)$ and $\hat{x}_{ij}(t)$ as time evolves and new measurements become available. 
\section{Filter Derivation}\label{filterder} 
In the following we proceed by substituting the measurements and
approximation models into the cost. Substituting
equations~\eqref{measurementi},~\eqref{measurementsj} and~\eqref{belief}
into the cost~\eqref{cost} yields 
\begin{equation}\label{rcost}
\begin{split}
 &J_{i,t}(x_i(0),\delta_i,\{\eta_{ij}\}) =
 \frac{\Xi_i}{2} (x_i(0)-x_{i0} )^ 2\\
 &+ \frac{1}{2} \int_0^t [ \delta_i^2+\frac{ (y_{ii}-x_i)^2}{R_{ii}}\\
 &+ \sum_{j\in\mathcal{N}_i}(\frac{( y_{ij}-x_i-\eta_{ij})^2}{R_{ij}} +  \frac{\eta_{ij}^2}{G_{ij}})]d\tau,
 \end{split}
 \end{equation}
 where the weightings are denoted by $R_{ii}\coloneqq D_{ii}^2$ and $R_{ij}\coloneqq D_{ij}^2$. 
 
 Therefore, in order to solve the minimum-energy filtering problem we need to solve the following two step optimization problem,
 \begin{equation}\label{prob}
  \inf_{x_i(0)}(\inf_{\delta_i,\;\{\eta_{ij}\}} J_{i,t}(x_i(0),\delta_i,\{\eta_{ij}\})).
 \end{equation}
 
 We solve the inner problem by minimizing the cost~\eqref{rcost} over the approximation errors $\{\eta_{ij}\}$ as if they are control signals similar an optimal control problem~\cite{Athans}. The minimizing values are
 \begin{equation}\label{optdisagr}
 \begin{split}
 &\eta_{ij}^* = \frac{G_{ij}}{(G_{ij}+R_{ij})}(y_{ij}-x_i).
 \end{split}
 \end{equation}
Replacing~\eqref{optdisagr} into the cost and using the matrix inversion lemma yields
\begin{equation}\label{orcost}
\begin{split}
 &J_{i,t}(x_i(0),\delta_i,\{\eta_{ij}\}) =
 \frac{\Xi_i}{2} (x_i(0)-x_{i0} )^ 2\\
 &+ \frac{1}{2} \int_0^t [ \delta_i^2+\frac{ (y_{ii}-x_i)^2}{R_{ii}}+ \sum_{j\in\mathcal{N}_i}\frac{( y_{ij}-x_i)^2}{S_{ij}} ]d\tau,
 \end{split}
 \end{equation}
where  $S_{ij}$ is defined as
\begin{equation}
 S_{ij}\coloneqq R_{ij}+G_{ij}.
\end{equation}

 Using the cost functional~\eqref{orcost} and applying minimum-energy filtering~\cite{Mortensen,Hijab} to solve problem~\eqref{prob} yields the following local minimium-energy filter. The observer equation of the filter at node $i$ is 
 \begin{equation}\label{obs}
 \dot{\hat{x}}_i = \sum_{j\in\mathcal{N}_i}(\hat{x}_{ij} - \hat{x}_i) + Q_i[\frac{y_{ii}-\hat{x}_i}{R_{ii}}+\sum_{j\in\mathcal{N}_i}\frac{y_{ij}-\hat{x}_i}{S_{ij}}],
\end{equation}
where $\hat{x}_i(0)=x_{i0}$.
 The neighboring estimates at node $i$, $\hat{x}_{ij}$, are computed statically
\begin{equation}\label{optxhatj}
 \hat{x}_{ij} =  \hat{x}_i + \frac{G_{ij}}{S_{ij}}(y_{ij}-\hat{x}_i).
\end{equation}
The observer gain  $Q_i$ is required for the observer~\eqref{obs} and is obtained from the Riccati equation:
\begin{equation}\label{iRiccati}
 \dot{Q}_i = B_i^2-Q^2_i(\frac{1}{R_{ii}}+\sum_{j\in\mathcal{N}_i}\frac{1}{S_{ij}}),\;Q_i(0)=\frac{1}{\Xi_i}.
\end{equation}
\section{Convergence Results}\label{convergence} 
 In this section we focus on convergence properties of the proposed
 consensus algorithm that was derived in the previous section.
  
Before continuing any further we consider how $Q_i$ evolves in Equation~\eqref{iRiccati}. The evolution of $Q_i$ is independent of the rest of the algorithm and it can be seen that it converges to $Q_i^*=B_i^2(\frac{1}{R}+\sum_{j\in\mathcal{N}_i}\frac{1}{S})^{-\frac{1}{2}}$, $i=1,\dots,N$ for any choice of $\Xi_i$. 
Thus one can simply pick 
\begin{equation}\label{eq:init_cond_ricc}
\Xi_i=(Q_i^*)^{-1}.
\end{equation} This makes $Q_i=Q_i^*$ independent of time and renders our consensus algorithm time-invariant. The corresponding consensus algorithm under this initialisation is summarized as follows. 
\begin{equation}\label{algorithm}
\boxed{
 \begin{split}
   &\dot{x}_i =  \sum_{j\in\mathcal{N}_i}(\hat{x}_{ij} - \hat{x}_i) +B_i\delta_i,\\
   &\dot{\hat{x}}_i = \sum_{j\in\mathcal{N}_i}(\hat{x}_{ij} - \hat{x}_i) + Q^*_i[\frac{y_{ii}-\hat{x}_i}{R_{ii}}\\
   &\hspace{25mm}+\sum_{j\in\mathcal{N}_i}\frac{y_{ij}-\hat{x}_i}{S_{ij}}],\quad
   \hat{x}_i(0)=x_{i0}\\
   &\hat{x}_{ij}  = \hat{x}_i +\frac{G_{ij}}{S_{ij}}(y_{ij}-\hat{x}_i),\\
  \end{split}}
 \end{equation}
For ease of notation and computation we assume that all the agents have equal tuning parameters
 \begin{equation}\label{samep}
 \begin{split}
R_{ii}=R,\quad
 S_{ij}=S,\quad
 G_{ij}=1.
 \end{split}
 \end{equation}
This assumption is the case for instance in a network where all the nodes utilize the same kind of sensors and have access to same quality communication links. 
  
First, convergence of the disturbance-free algorithm is analyzed. 
Consider the local estimation error at node $i$  defined as $e_i\triangleq \hat{x}_i-x_i$. Hence and by using~\eqref{algorithm},~\eqref{measurementi}, ~\eqref{measurementsj} and~\eqref{samep}, the local error system in the absence of the disturbances $\epsilon_{ii}$, $\delta_i$, and $\{\epsilon_{ij}\}$ can be written as
\begin{equation}\label{errors}
\begin{split}
&\dot{x}_i = \frac{1}{S}\sum_{j\in\mathcal{N}_i} (x_j-x_i-e_i),\\
&\dot{e}_i =  Q^*_i[-\frac{e_i}{R}+\frac{1}{S}\sum_{j\in\mathcal{N}_i}(x_j-x_i-e_i)],\\
\end{split}
\end{equation}
Consider the following notations regarding the parameters of the overall network, that are going to be used in our analysis. 
\begin{equation}\label{parameters}
 \begin{split}
  Q^*\triangleq \mbox{Diag}(Q^*_i),\quad  
  \tilde{L} \triangleq \frac{1}{S}L ,\quad
  \tilde{\Delta} \triangleq \frac{1}{S}\Delta .
 \end{split}
\end{equation}
The Laplacian $L$ and the degree matrix $\Delta$ are as defined in Section~\ref{consensus}.

Denote the global estimation error by $e\triangleq [e_1,\cdots,e_N]^{\top}$ and denote the global state by $x\triangleq [x_1,\cdots, x_N]^{\top}$.  The global system satisfies the following dynamics.
\begin{equation}\label{globalsys}
  \left[\begin{array}{c}
         \dot{x}\\ \dot{e}
        \end{array}\right] = F\left[\begin{array}{c}
         x\\ e
        \end{array}\right],\quad F\triangleq \left[\begin{array}{cc}\tilde{L} & -\tilde{\Delta}\\Q^*\tilde{L} & -Q^*(\frac{1}{R}I +\tilde{\Delta})\end{array}\right].
 \end{equation}
%
Next we will state the stability properties of~\eqref{globalsys}.

 \begin{theorem}\label{th:convergence}
Suppose that the assumptions~\eqref{eq:init_cond_ricc} and \eqref{samep} hold and the network is strongly connected. 
Then, the global system~\eqref{globalsys} converges to $x_1=x_2=\cdots=x_N=x^*$  and $e^*=0$ exponentially fast, where 
\begin{equation}\label{eq:eqlb_point}
x^* =\dfrac{ \omega^\top (I +R\tilde{\Delta}) x(0)-\omega^\top (R\Xi\tilde{\Delta}) e(0)}{ \omega^\top (I +R\tilde{\Delta}) \mathbf{1}},
\end{equation}
here $\omega$ is the left eigenvector of $\tilde{L}$ corresponding to its zero
eigenvalue, i.e., $\omega^\top \tilde{L}=0$, and $\Xi \triangleq
\mbox{Diag}(\Xi_i)$. 
\end{theorem}

 The proof of Theorem~\ref{th:convergence} is based on the spectral
properties of $F$ established in Lemma~\ref{Lem:constant} below.

\begin{lemma}\label{Lem:constant}
Let $q$ be the geometric multiplicity of the zero eigenvalue of $L$. 
The matrix $F$ has $2N-q$ eigenvalues with negative
real parts, and the geometric multiplicity of the zero eigenvalue of $F$ is
equal to $q$.
\end{lemma}
The proof is a practice in algebra but nevertheless standard in spectral analysis of networked systems.

\begin{remark}
If the network is balanced, i.e.,~the in-degree and out-degree of each node are
equal, then $\omega =\mathbf{1}$ (e.g.,~see Theorem 6 of
\cite{olfati2004consensus}) and 
$$
x^* =\dfrac{ (\mathbf{1} +R\tilde{\Delta}\mathbf{1})^\top x(0)-(R\Xi\tilde{\Delta}\mathbf{1})^\top e(0)}{N+(R/S)\sum_{i=1}^N d_i},
$$
where $d_i$ is the degree of node $i$ as introduced in Section \ref{consensus}. 
\end{remark}

  Theorem~\ref{th:convergence} demonstrated that the algorithm converges to $[x^*\mathbf{1}^{\top}\; \mathbf{0}^{\top}]^{\top}$  in the absence of any disturbance exponentially fast and there exists $a>0, b>0$ such that for all $0 < t <\infty$
\begin{equation}\label{eq:exp_ball}
\left \| \begin{bmatrix} x(t)\\ e(t)\end{bmatrix} \right \| \leq b \left \| \begin{bmatrix} x(0)\\ e(0)\end{bmatrix} \right \| e^{-a t}.
\end{equation} 
In what follows, we demonstrate the performance of the algorithm in the presence of bounded and continuous disturbances. We have the following assumption for the disturbances:
\begin{assumption}\label{asmp:bounded_dist}
Let $\delta_i(t)$, $i=1,\dots,N$ and $\epsilon_{ij}(t)$, $(i,j)\in\mathbf{E}$ be bounded and continuous on $[0, \infty)$ such that $|\delta_i(t)|\leq \delta_{max}$ and $|\epsilon_{ij}(t)|\leq \epsilon_{max}$.
\end{assumption}

In the presence of disturbances, the system and the error dynamics~\eqref{errors} becomes
\begin{equation}\label{errors_nonzero_dist}
\begin{split}
&\dot{x}_i = \frac{1}{S}\sum_{j\in\mathcal{N}_i} (x_j-x_i-e_i)+\vartheta_i ,\\
&\dot{e}_i =  Q^*_i[-\frac{e_i}{R}+\frac{1}{S}\sum_{j\in\mathcal{N}_i}(x_j-x_i-e_i)] + \zeta_i,\\
\end{split}
\end{equation}
where 
\begin{equation}\label{eq:pert_dist_i}
\begin{split}
\vartheta_i &= \frac{1}{S}\sum_{j\in\mathcal{N}_i}  \epsilon_{ij}+B_i \delta_i,\\
\zeta_i &= Q^*_i[\frac{\epsilon_{ii}}{R}+\frac{1}{S}\sum_{j\in\mathcal{N}_i}  \epsilon_{ij}].
\end{split}
\end{equation}
We have the following theorem.
\begin{theorem}\label{lem:cons_pert}
Consider Assumption \ref{asmp:bounded_dist}, \eqref{eq:init_cond_ricc}, \eqref{samep} and the system described in~\eqref{errors_nonzero_dist} for a strongly connected network. Then for $0 <t<\infty$ and $a$ and $b$ defined in~\eqref{eq:exp_ball}  we have
\begin{equation}
\left \| \begin{bmatrix} x(t)\\ e(t)\end{bmatrix} \right \|  \leq b \left \| \begin{bmatrix} x(0)\\ e(0)\end{bmatrix} \right \| e^{-a t}+\frac{b \varphi_{max}}{a} \big( 1-
e^{-a t} \big), \label{Dangelis_pert}
\end{equation}
where 
\begin{equation}\label{eq:dist_bound}
\begin{split}
\varphi_{max} &= \epsilon_{max} \frac{\sum_{i=1}^N N_i}{S} +\delta_{max}\sum_{i=1}^N B_i\\
&+Q_{max}[\frac{N\epsilon_{max}}{R}+\frac{\epsilon_{max}}{S}\sum_{i=1}^N  N_i],
\end{split}
\end{equation}
where $Q^*_i \leq Q_{max}$, for all $i=1,\dots,N$, and $N_i=|\mathcal{N}_i|$.
\end{theorem}
\begin{remark}
The upper bound on the magnitude of the disturbance $\varphi$ given in~\eqref{eq:dist_bound} is not unique and other bounds can be proposed. 
\end{remark}
Theorem~\ref{lem:cons_pert} immediately establishes an ISS property of the trajectories of the system under continuous and bounded disturbances.

We conclude this section by commenting briefly on the impact of values of $R$ and $B_i$ on the proposed algorithm. First, we study the impact of $R$. From \eqref{eq:eqlb_point} it can be seen that as $R\rightarrow 0$, $x^*\rightarrow \omega^\top x(0)/(\omega^\top \mathbf{1})$ which is the solution to the classical consensus problem without disturbances. However, for very small values of $R$ the upper bound $\varphi_{max}$ becomes very large which shows that achieving accuracy and robustness are at odds. Next, we consider the impact of $B_i$. Large values of $B_i$ makes $\Xi$ very small and reduces the impact of the initial errors of the filters on $x^*$, however, as in the case of $R$ it has an adverse impact on $\varphi_{max}$.

It is interesting to point out that the dichotomy of achieving accuracy and robustness in reaching consensus has been encountered  in some other existing results, even though the nature of the proposed algorithms are very different, e.g.~see \cite{sundaram2011distributed}.

\section{Numerical Examples}\label{numexp}
\begin{figure}[htbp] 
   \centering
   \includegraphics[width=8.5cm]{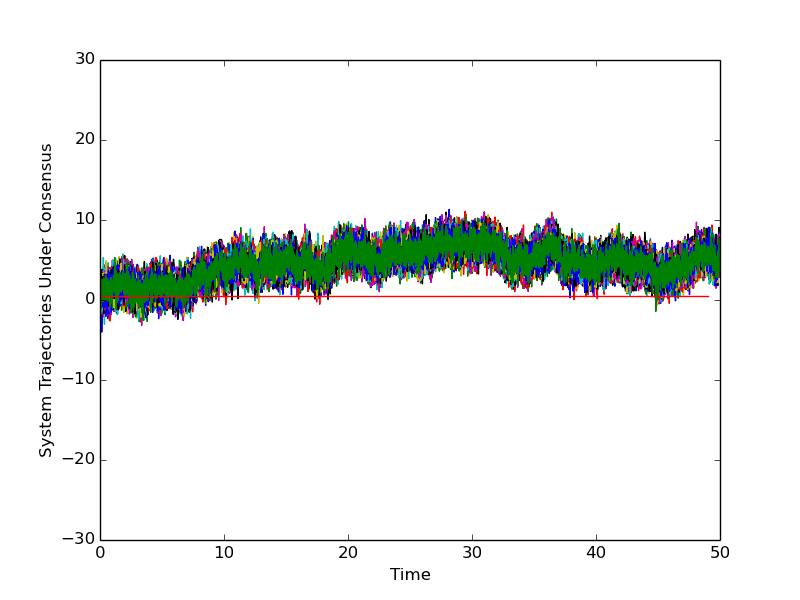} 
   \caption{The trajectories of 100 systems under the consensus algorithm \eqref{eq:cons} in the presence of stochastic disturbances.}
   \label{fig:cons}
\end{figure}

\begin{figure}[htbp] 
   \centering
   \includegraphics[width=8.5cm]{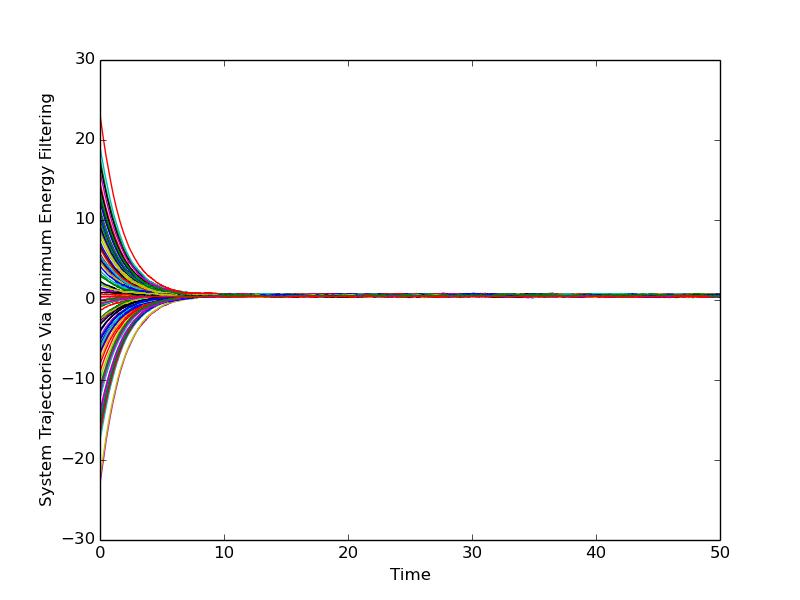}   \caption{The trajectories of 100 systems under minimum energy filtering based algorithm \eqref{algorithm} in the presence of stochastic disturbances.}
   \label{fig:mec}
\end{figure}

In this section, we consider the problem of \emph{coherence} in networks. To this aim consider a network governed by 
\begin{equation}\label{eq:cons}
\dot{x}= -L x + \delta
\end{equation}
where $x\in\mathbb{R}^N$ and $\delta\in\mathbb{R}^N$ is a zero-mean mutually independent white stochastic process. It is known that the sum of expected deviation of the nodes' states from the average, $\mathcal{D}_{ave}$, for the network satisfies:
\begin{equation}
\begin{split}
\mathcal{D}_{ave}&= \lim_{t\rightarrow \infty} \sum_{i=1}^N  \mathbb{E} \Big[\big (x_i -\dfrac{1}{n}\sum_{j=1}^N x_j \big )^2 \Big ]\\
&= \dfrac{1}{2} \sum_{i=2}^N \dfrac{1}{\lambda_i(L)}
\end{split}
\end{equation}
where $\mathbb{E}[\cdot]$ is the expectation of its argument and $0=\lambda_1(L)<\lambda_2(L)\leq \dots\leq \lambda_N(L)$ are the eigenvalues of $L$. For more information the reader may refer to \cite{bamieh2012coherence} and \cite{young2011rearranging}. It is immediate that  the sum of expected deviation of the nodes' states from the average  for a fixed number of nodes is minimised if the network is isomorphic to a complete graph as its nonzero eigenvalues take the maximum possible value, i.e., $\lambda_i(L)=N$, $i=2,\dots,N$. As a best case scenario we consider such a network. We depict the state values for a complete graph of $N=100$ nodes in Fig.~\ref{fig:cons}. On the other hand the proposed  algorithm~\eqref{algorithm} is implemented for the same network with $R=1$,  $S=1$, and $B_i=1$ for all $i=1,\dots,N$ under the same disturbance levels and initial conditions as Fig.~\ref{fig:cons}. The trajectories of the systems for \eqref{algorithm} are presented in Fig.~\ref{fig:mec}. The proposed algorithm clearly outperforms the consensus algorithm~\eqref{eq:cons} even in the most favourable interconnection for \eqref{eq:cons}.

%
%
%
%
%
%
%
%
%
\section{Conclusions}\label{conc}  
 In this paper we considered the problem of reaching consensus in the presence of disturbances in measurements and inputs. We proposed an algorithm based on minimum energy filtering without making an assumption on the properties of the disturbances. We showed that the proposed algorithm converges to the solution of the traditional consensus algorithm in the disturbance free case. Moreover, we studied the algorithm under a class of bounded and continuous disturbances. Furthermore, the performance of the proposed  algorithm was demonstrated via a numerical example. An immediate future research direction is to explicitly show how the performance of the algorithm varies for other types of disturbances, and thoroughly study the impact of the algorithm parameters on its performance in the terms of accuracy, robustness, and the converge rate and propose a systematic way of choosing the parameters for different requirements.

\bibliographystyle{plain}

\bibliography{ref}
 
\end{document}